\documentclass{article}

\usepackage{arxiv}

\usepackage[utf8]{inputenc} 
\usepackage[T1]{fontenc}    
\usepackage{hyperref}       
\usepackage{url}            
\usepackage{booktabs}       
\usepackage{multirow}
\usepackage{amsmath}
\usepackage{amssymb}
\usepackage{amsfonts}       
\usepackage{nicefrac}       
\usepackage{microtype}      
\usepackage{cleveref}       
\usepackage{graphicx}
\usepackage{natbib}
\usepackage{doi}

\title{Not All EEG Moments Are Equal: Position-Adaptive Time Scheduling for EEG Generation}
\date{}
\newif\ifuniqueAffiliation
\uniqueAffiliationtrue

\author{
	Boheng Liu \\
	School of Computer Science and Technology\\
	Beijing Institute of Technology\\
	Beijing, China \\
	\texttt{boheng@bit.edu.cn} \\
	\And
	Ziyu Li\thanks{Corresponding author.} \\
	School of Computer Science and Technology\\
	Beijing Institute of Technology\\
	Beijing, China \\
	\texttt{ziyuli@bit.edu.cn} \\
	\And
	Chenghua Duan \\
	School of Computer Science and Technology\\
	Beijing Institute of Technology\\
	Beijing, China \\
	\And
	Qing Li \\
	School of Computer Science and Technology\\
	Beijing Institute of Technology\\
	Beijing, China \\
	\And
	Xia Wu \\
	School of Computer Science and Technology\\
	Beijing Institute of Technology\\
	Beijing, China \\
}

\renewcommand{\shorttitle}{Position-Adaptive Time Scheduling for EEG Generation}

\hypersetup{
pdftitle={Not All EEG Moments Are Equal: Position-Adaptive Time Scheduling for EEG Generation},
pdfsubject={cs.LG, eess.SP},
pdfauthor={Boheng Liu, Ziyu Li, Chenghua Duan, Qing Li, Xia Wu},
pdfkeywords={EEG generation, flow matching, brain-computer interface, data augmentation},
}

\begin{document}
\maketitle

\begin{abstract}
Electroencephalography (EEG) generation is essential for alleviating data scarcity and enabling large scale neural modeling in brain computer interface applications.
However, existing flow based approaches assume that every channel and every time segment within a sample shares a single global time progression, overlooking the fact that not all EEG moments are equal.
To address this overlooked heterogeneity, we propose an adaptive EEG generation framework built on conditional flow matching.
The framework introduces Position-Adaptive Time Scheduling, which tracks per position reconstruction error to modulate a position specific time progress within the flow matching trajectory.
It further incorporates Factorized Spatio-Temporal Attention and a frequency aligned multi resolution spectral consistency loss to model inter channel dependencies induced by volume conduction and compensate for the power law spectral bias of EEG, thereby improving the quality of generated signals.
Extensive experiments on three EEG datasets with distinct acquisition protocols and task semantics show that our framework consistently outperforms the strongest baseline, reducing TS-FID by up to 62.2\% and improving downstream classification accuracy gain by up to 6.77 percentage points.
These results suggest that the proposed method represents a promising step toward scalable, high fidelity data augmentation for real world brain computer interface applications.
\end{abstract}


\section{Introduction}

Electroencephalography (EEG) records the electrical activity of the brain with high temporal resolution, capturing rhythmic oscillations, transient events, and pathological patterns \citep{vanede2018neural}.
It plays a central role in brain computer interfaces and neuroscience research, with broad applications across numerous fields, such as sleep staging \citep{perslev2021usleep}, seizure detection \citep{yogarajan2023binary}, motor imagery decoding \citep{lawhern2018eegnet}, and emotion recognition \citep{zheng2019stable}.
At present, high quality EEG data serves as a critical data source for the brain computer interface field \citep{yang2025multiday}.
However, EEG data collection is constrained by the bottleneck of data scarcity, due to the high acquisition cost and the difficulty of expert annotation, which in turn limits performance across multiple downstream tasks \citep{you2025virtualeeg}.
Generative modeling offers a promising avenue to alleviate this bottleneck, since synthetic EEG can augment under represented pathological categories, while substantially reducing the cost of data acquisition \citep{vetter2024generating}.
Therefore, high quality EEG data generation is not merely a benchmark task, it represents an important direction for extending data driven neural modeling to clinically meaningful brain computer interface applications.

Early methods adopted generative adversarial networks to synthesize EEG signals for downstream augmentation, but such approaches, including representative works like ESC-GAN, remain prone to unstable training despite dedicated losses for minority category augmentation \citep{hartmann2018eeggan,zhang2025escgan}.
Subsequently, diffusion based frameworks, such as Diffusion-TS and PaD-TS, were introduced into general time series modeling, improving generation fidelity through clean sample reconstruction and population level statistical constraints \citep{vetter2024generating,yuan2024diffusionts,li2024padts}.
Despite these advances, both GAN based and diffusion based methods rely on discrete generative formulations, which are misaligned with the continuous nature of neural dynamics.
To address this mismatch, JET reformulates EEG generation as conditional flow matching, learning a continuous vector field that transports noise to the EEG data distribution \citep{wang2026jet}.
This continuous formulation substantially improves the fidelity of prior discrete paradigms, yet it still assumes that every channel and every time segment within a sample shares a single global time progression, overlooking the heterogeneous denoising difficulty across channels and time inherent to EEG data, which in turn limits generation performance.

\begin{figure}[t]
\centering
\includegraphics[width=0.9\linewidth]{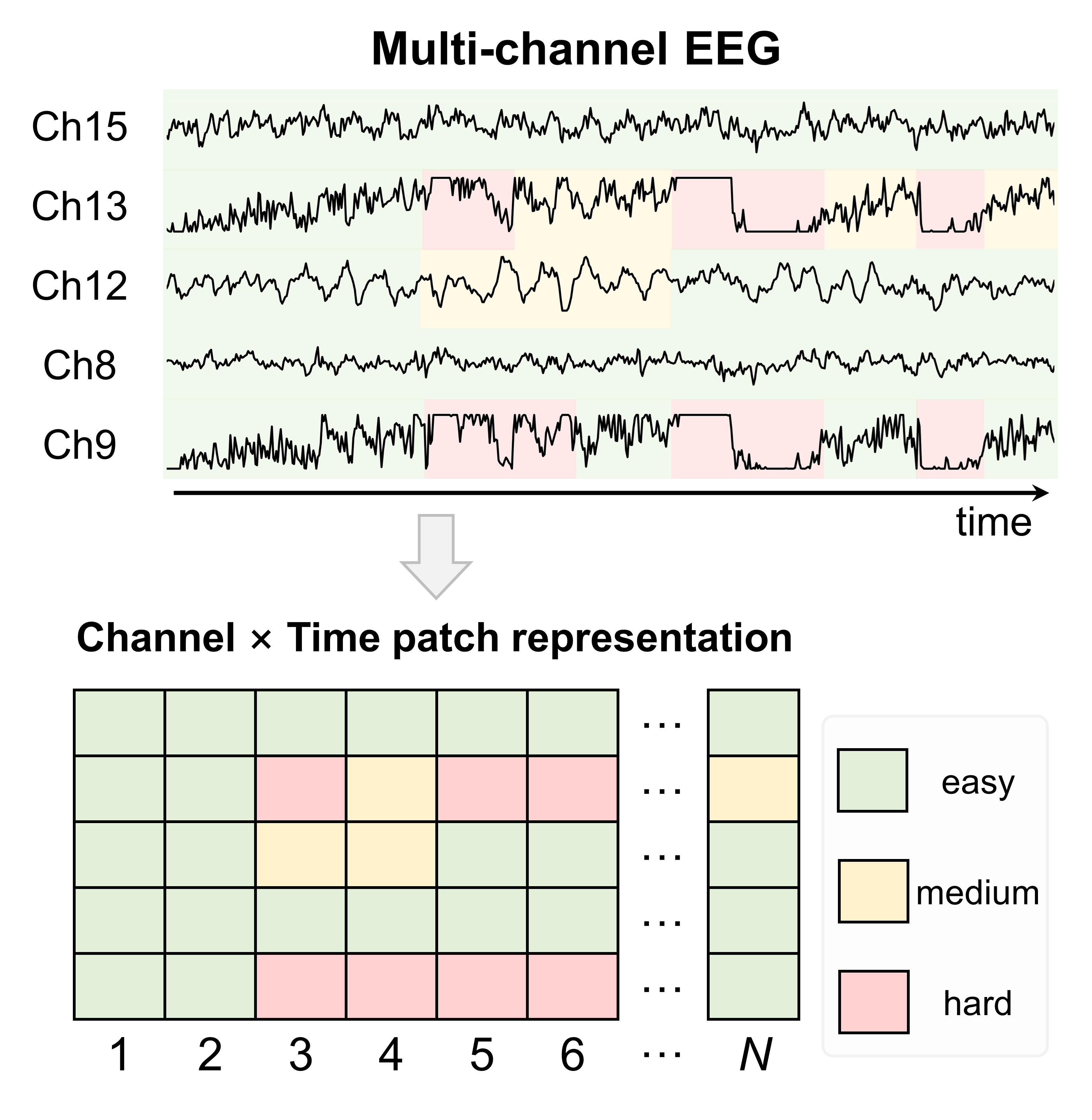}
\caption{Illustration of the overlooked heterogeneity in EEG generation.
Quasi stationary background activity (green) coexists with medium (yellow) and high difficulty (red) transient segments at different channel-time positions. Such patterns tend to recur at similar positions across recordings under the same protocol, motivating a position-wise difficulty statistic aggregated over the dataset rather than per instance.}
\label{fig:motivation}
\end{figure}

The prevailing assumption of uniform time scheduling contradicts a fundamental property of EEG signals, since not all EEG moments are equal, as illustrated in Figure~\ref{fig:motivation}.
Within a single recording, quasi stationary background rhythms coexist with sparse, high difficulty transients, such as pathological discharges, artifact contaminated segments, and channel specific physiological events, which tend to recur at similar channel-time positions across a given acquisition protocol; yet a shared linear interpolation path forces every channel and patch position to traverse the same interpolation progress at the same training step, regardless of how consistently difficult that position has proven to be across the dataset \citep{vanede2018neural,lipman2023flowmatching}.
As a result, easy and difficult regions receive identical effective supervision, preventing the limited model capacity from being fully utilized.
We identify this overlooked heterogeneity as the central bottleneck limiting the fidelity of EEG generation, motivating the key research question of this work: how a generative model can adaptively target complex segments according to the recurring difficulty structure observed during training to efficiently improve generation quality.

To address the above problem, we propose an EEG generation framework capable of learning the heterogeneity of denoising difficulty across channels and time, with three main contributions.
First, Position-Adaptive Time Scheduling (PATS) maintains an exponential moving average of the reconstruction error at each channel and patch position, modulating the position specific time progress to allocate training signal according to the historically observed difficulty at that position.
Second, a Factorized Spatio-Temporal Attention module decomposes global self attention into temporal and channel stages conditioned on this schedule, explicitly modeling inter channel dependencies induced by volume conduction.
Third, a frequency aligned multi resolution spectral consistency loss replaces the time domain smoothness constraint, reweighting frequency bands to compensate for the power law spectral bias of EEG.

Extensive experiments on three EEG datasets with distinct acquisition protocols and task semantics show that our framework consistently outperforms the strongest baseline, reducing TS-FID by up to 62.2\% and improving the downstream accuracy gain by up to 6.77 percentage points over the strongest baseline.
These results suggest that position-adaptive generative modeling holds promise for broader physiological time series applications, pointing toward a general paradigm for data augmentation in data-scarce, high-stakes brain computer interface scenarios.

\begin{figure*}[h]
\centering
\includegraphics[width=\textwidth]{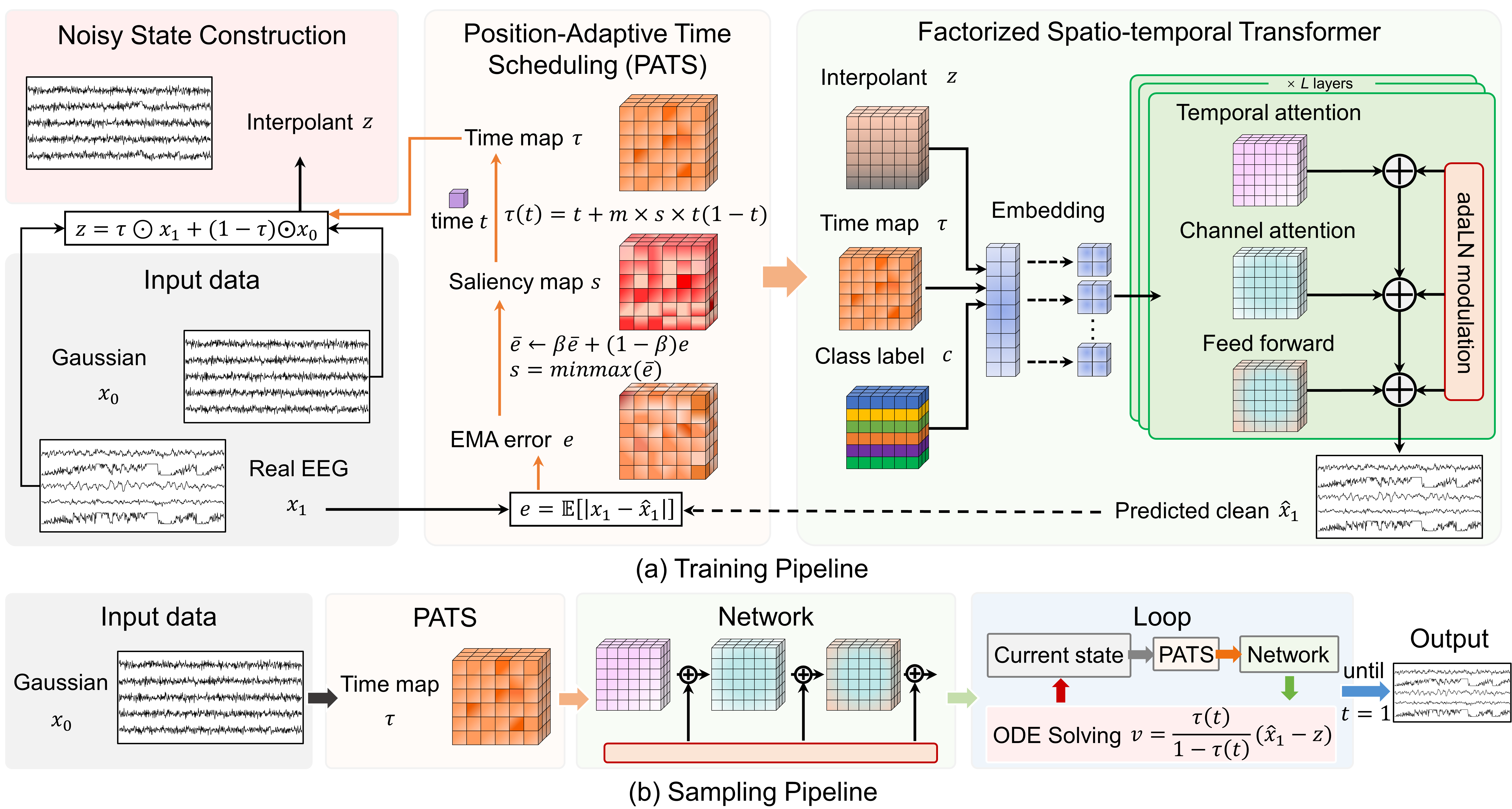}
\caption{Overview of the proposed framework. (a) Training pipeline: PATS derives a position specific time map from the EMA error map, which conditions the Factorized Spatio-temporal Transformer to predict the clean sample. (b) Sampling pipeline: starting from Gaussian noise, PATS and the network are applied iteratively to solve the flow ODE and generate the final EEG output.}
\label{fig:framework}
\end{figure*}

\section{Related Work}

\subsection{Generative Modeling for Data Synthesis}

Generative modeling has evolved rapidly over the past decade, reshaping data synthesis across numerous domains.
GANs pioneered this era via adversarial training, yet GANs remain prone to mode collapse on imbalanced or high dimensional data \citep{goodfellow2014gan,gulrajani2017wgangp,karras2019stylegan}.
Diffusion and score based models emerged as more stable alternatives via iterative denoising, extended to sequential data through diffusion forecasting and imputation \citep{ho2020ddpm,song2021scorebased,rasul2021timegrad,tashiro2021csdi}, yet their discrete trajectories misalign with continuously evolving data.
Flow matching and rectified flow instead learn a continuous vector field transporting a prior to the data distribution, enabling straighter, more efficient sampling, and combined with scalable backbones such as DiT and JiT have achieved strong fidelity across image and sequential modeling \citep{lipman2023flowmatching,peebles2023dit,li2025jit}.

\subsection{EEG Signal Generation}

Research on EEG signal generation has followed a similar trajectory, beginning with GAN based approaches.
EEG-GAN adapts adversarial training to synthesize raw EEG waveforms, subsequent works explored synthetic EEG for BCI augmentation and spatial super-resolution, and ESC-GAN edits reference signals from well represented to under represented emotional subspaces through diversity aware and boundary aware losses \citep{hartmann2018eeggan,fahimi2021ganbci,zhang2025escgan}.
However, GAN based EEG generators remain limited by unstable adversarial training and restricted mode coverage, hindering their ability to capture the full variability of real recordings.
Diffusion based methods were subsequently explored to improve generation stability, including EEG specific frameworks for seizure prediction and ERP synthesis, as well as general models such as Diffusion-TS and PaD-TS adapted to physiological signals \citep{shu2024diffeeg,yuan2024diffusionts,li2024padts}, yet these still rely on discretized denoising schedules that treat every time step and channel uniformly, conflicting with the continuous, non-stationary evolution of neural activity.
To resolve this mismatch, JET reformulates EEG generation as conditional flow matching with principled constraints on spectral, temporal, and statistical structure, achieving state of the art fidelity among existing methods \citep{wang2026jet}.
Nevertheless, JET still assumes a single global time progression shared across all channels and time segments, leaving the position dependent nature of EEG signals insufficiently modeled and constraining further improvement.

\section{Method}

\subsection{Preliminaries}

Let $\mathbf{X} \in \mathbb{R}^{C \times T}$ denote a raw multi-channel EEG segment, where $C$ is the number of channels and $T$ is the number of time points.
We divide the temporal axis into $N = T / P$ non-overlapping patches of length $P$, yielding a patchified representation $\mathbf{x}_1 \in \mathbb{R}^{C \times N \times P}$, a clean sample from the EEG data distribution $q(\mathbf{x}_1)$, and let $\mathbf{x}_0 \sim \mathcal{N}(\mathbf{0}, \mathbf{I})$ denote a sample from the Gaussian prior with the same shape.

Given a pair $(\mathbf{x}_0, \mathbf{x}_1)$ and a global time $t \in [0,1]$, conditional flow matching defines the linear interpolation path $\mathbf{x}_t = t \mathbf{x}_1 + (1-t) \mathbf{x}_0$, whose time derivative yields the target vector field $\mathbf{u}_t(\mathbf{x}_t \mid \mathbf{x}_0, \mathbf{x}_1) = \mathbf{x}_1 - \mathbf{x}_0$.
A neural network $f_\theta$, conditioned on the noisy state $\mathbf{x}_t$, the time $t$, and the pathological class label $c$, is trained to predict the clean sample $\hat{\mathbf{x}}_1 = f_\theta(\mathbf{x}_t, t, c)$, from which the vector field estimate is recovered as $\mathbf{v}_\theta(\mathbf{x}_t, t, c) = (\hat{\mathbf{x}}_1 - \mathbf{x}_t) / (1-t)$.
During inference, samples are generated by numerically solving $d\mathbf{x}_t / dt = \mathbf{v}_\theta(\mathbf{x}_t, t, c)$ starting from $\mathbf{x}_0$.
This formulation assumes a single scalar $t$ shared by every channel and patch position, the uniform scheduling assumption we relax below.

\subsection{Overall Framework}

Our framework combines a difficulty aware time scheduler, which expands the global time $t$ into an endpoint preserving, position specific time map, with a factorized spatio-temporal transformer that consumes this map as a token level condition, as illustrated in Figure~\ref{fig:framework}.

\subsubsection{Position-Adaptive Time Scheduling}

Recurring transients and artifact contaminated segments manifest during training as positions with systematically larger reconstruction error, making that error a practical proxy for position dependent difficulty.
To allocate training signal by position, we maintain a difficulty statistic for every channel and patch throughout training.
Let $\bar{\mathbf{e}} \in \mathbb{R}^{C \times N}$ denote the exponential moving average of the per position reconstruction error, updated after every training step as $\bar{\mathbf{e}} \leftarrow \beta \bar{\mathbf{e}} + (1-\beta) \mathbf{e}$, where $\mathbf{e}_{i,j} = \mathbb{E}_{P,\mathcal{B}}[\lvert \mathbf{x}_{1,i,j} - \hat{\mathbf{x}}_{1,i,j} \rvert]$ is the mean absolute error at position $(i,j)$ averaged over the patch dimension $P$ and the batch $\mathcal{B}$, and $\beta \in (0,1)$ is the decay rate.
This statistic is shared across all samples in the dataset and updated continuously throughout training, reflecting the difficulty a given channel-time position has exhibited on average rather than a property estimated per instance.
We normalize $\bar{\mathbf{e}}$ into a saliency map $\mathbf{s} \in [0,1]^{C \times N}$ via min-max scaling, where higher values indicate positions that have historically been harder to reconstruct.

Any position specific reparameterization of the flow matching trajectory must preserve the endpoints of the interpolation path, since the generative process starts from the Gaussian prior $\mathbf{x}_0$ at $t=0$ and reaches the data distribution at $t=1$ regardless of position.
Given a global time $t$, we expand it into a position specific time map $\boldsymbol{\tau} \in [0,1]^{C \times N}$ as
\begin{equation}
\tau_{i,j}(t) = t + m \, s_{i,j} \, t(1-t),
\end{equation}
where $m \in [0,1)$ is a margin coefficient controlling the maximum deviation from the uniform schedule, and $m=0$ recovers the original schedule.
This form satisfies $\tau_{i,j}(0)=0$ and $\tau_{i,j}(1)=1$ for every position regardless of $s_{i,j}$, so every position still starts from pure noise and ends at the clean sample, with only the intermediate progress speed modulated by historical difficulty; it further remains strictly monotonic in $t$ for $m<1$, since $\dot{\tau}_{i,j}(t) = 1 + m\,s_{i,j}(1-2t) > 0$ over $[0,1]$.
Positions with higher historical difficulty are thus advanced toward a less noisy stage earlier than the global schedule.

The noisy state is constructed at the patch level as $\mathbf{z} = \boldsymbol{\tau} \odot \mathbf{x}_1 + (1-\boldsymbol{\tau}) \odot \mathbf{x}_0$, with $\boldsymbol{\tau}$ broadcast along the patch dimension $P$, and the network is reformulated to take this time map in place of the scalar $t$:
\begin{equation}
\hat{\mathbf{x}}_1 = f_\theta(\mathbf{z}, \boldsymbol{\tau}, c).
\end{equation}

Differentiating the interpolation path through $\tau_{i,j}(t)$ and replacing $x_{1,i,j}$ with the network prediction $\hat{x}_{1,i,j}$ gives the position specific vector field used for sampling,
\begin{equation}
v_{i,j}(\mathbf{z}, t, c) = \frac{\dot{\tau}_{i,j}(t)}{1-\tau_{i,j}(t)}\,\big(\hat{x}_{1,i,j} - z_{i,j}\big),
\end{equation}
with $\dot{\tau}_{i,j}(t) = 1 + m\,s_{i,j}(1-2t)$, differing from a scalar schedule by this position dependent factor, whose omission would implicitly integrate the ODE as if $m=0$.

At generation time, the same statistic $\bar{\mathbf{e}}$ from training is used to construct $\mathbf{s}$ and $\boldsymbol{\tau}$ at every sampling step, so the generation trajectory follows the same position specific schedule as training.
Starting from $\mathbf{z}$ sampled from the Gaussian prior at $t=0$, the ODE $d\mathbf{z}/dt = \mathbf{v}(\mathbf{z}, t, c)$ is solved over $t \in [0,1]$ with a Heun or Euler integrator, using Eq.~(3) at every step.

\subsubsection{Factorized Spatio-Temporal Attention}

Since $\boldsymbol{\tau}$ varies across both channels and patches, the conditioning signal must be resolved at the token level rather than the sample level.
Each patch of $\mathbf{z}$ is embedded into a $D$ dimensional token via linear projection, with learnable temporal position and channel embeddings added to obtain $\mathbf{h}^{(0)} \in \mathbb{R}^{C \times N \times D}$.
For each token at position $(i,j)$, a condition vector $\mathbf{g}_{i,j} = \mathrm{TimeEmbed}(\tau_{i,j}) + \mathrm{LabelEmbed}(c)$ predicts the shift, scale, and gate parameters of adaptive layer normalization at every block.

Rather than applying self attention jointly over all $C \times N$ tokens, each block factorizes attention into two sequential stages: a temporal stage attending over the $N$ patches within each channel to capture long range temporal dependencies, and a channel stage attending over the $C$ channels at each temporal position to explicitly model inter channel dependencies induced by volume conduction.
Both stages and the subsequent feed forward layer are modulated by the same token level condition $\mathbf{g}_{i,j}$, so the network perceives a different effective denoising progress at each channel-time position.
This factorization reduces the attention complexity from $\mathcal{O}((CN)^2)$ to $\mathcal{O}(CN^2 + NC^2)$ while preserving a global receptive field through the two sequential stages.
The final prediction $\hat{\mathbf{x}}_1$ is obtained by projecting the output tokens back to the patch size $P$ through a final layer conditioned on $\mathbf{g}_{i,j}$.

\subsection{Training Objective}

We train the network by supervising the predicted clean sample $\hat{\mathbf{x}}_1$ against the ground truth $\mathbf{x}_1$ under a composite training objective.

\subsubsection{Difficulty Weighted Reconstruction Loss}

Each position is assigned a weight derived from the saliency map $\mathbf{s}$ used in Position-Adaptive Time Scheduling, $w_{i,j} = 1 + \gamma \sqrt{s_{i,j}}$, where $\gamma \geq 0$ is a scaling coefficient ($\gamma=0$ recovers the unweighted mean), giving
\begin{equation}
\mathcal{L}_{recon} = \frac{\mathbb{E}_{i,j}\left[w_{i,j} \cdot \mathbb{E}_{P}\lvert \mathbf{x}_{1,i,j} - \hat{\mathbf{x}}_{1,i,j} \rvert\right]}{\mathbb{E}_{i,j}[w_{i,j}]},
\end{equation}
which emphasizes historically difficult positions while keeping the loss scale independent of $\gamma$.

\subsubsection{Statistical Consistency Loss}

To prevent drift in amplitude statistics, we match the per-channel temporal mean $\mu(\cdot)$ and standard deviation $\sigma(\cdot)$ between real and generated signals,
\begin{equation}
\mathcal{L}_{stat} = \lVert \mu(\mathbf{x}_1) - \mu(\hat{\mathbf{x}}_1) \rVert_1 + \lVert \sigma(\mathbf{x}_1) - \sigma(\hat{\mathbf{x}}_1) \rVert_1.
\end{equation}

\subsubsection{Frequency Aligned Multi-Resolution Spectral Consistency Loss}

Since EEG exhibits a power law spectral density that biases time domain smoothness constraints toward suppressing clinically relevant high frequency transients, we instead constrain the flow in the spectral domain via a multi-resolution STFT objective over $R$ resolutions.
Let $\mathbf{M}_x^{(r)}$ and $\mathbf{M}_{\hat{x}}^{(r)}$ denote the STFT magnitude spectra of $\mathbf{x}_1$ and $\hat{\mathbf{x}}_1$ at resolution $r$, with $F$ frequency bins indexed by $k$.
To prevent the dominant low frequency energy from suppressing the gradient at higher frequencies, we apply a frequency weight $\mathbf{w}_k \in \mathbb{R}^F$ with $w_k \propto k^{\alpha}$ ($\alpha \geq 0$) within the evaluated band $k < \eta F$ ($\eta \in (0,1]$ matching the downstream evaluation protocol's maximum frequency ratio), and $w_k=1$ outside it.
The per resolution loss combines a spectral convergence term and a log-magnitude term,
\begin{equation}
\begin{split}
\mathcal{L}_{spec}^{(r)} = \; & \frac{\lVert \mathbf{w}_k \odot (\mathbf{M}_x^{(r)} - \mathbf{M}_{\hat{x}}^{(r)}) \rVert_F}{\lVert \mathbf{w}_k \odot \mathbf{M}_x^{(r)} \rVert_F} \\
& + \big\lVert \mathbf{w}_k \odot \big[\log(\mathbf{M}_x^{(r)} + \delta) - \log(\mathbf{M}_{\hat{x}}^{(r)} + \delta)\big] \big\rVert_1,
\end{split}
\end{equation}
with $\delta$ a small constant for numerical stability, and the final spectral loss $\mathcal{L}_{spec} = \frac{1}{R}\sum_{r=1}^{R} \mathcal{L}_{spec}^{(r)}$ averages over resolutions to jointly capture transient and long term spectral structure.

\subsubsection{Correlation Loss}

To encourage waveform level alignment invariant to amplitude scaling, we maximize the Pearson correlation between the flattened real and generated signals,
\begin{equation}
\mathcal{L}_{corr} = 1 - \rho_{\mathrm{Pearson}}(\mathbf{x}_1, \hat{\mathbf{x}}_1).
\end{equation}

\subsubsection{Total Objective}

The final training objective combines the four terms above,
\begin{equation}
\mathcal{L}_{total} = \mathcal{L}_{recon} + \lambda_{stat} \mathcal{L}_{stat} + \lambda_{spec} \mathcal{L}_{spec} + \lambda_{corr} \mathcal{L}_{corr},
\end{equation}
where $\lambda_{stat}$, $\lambda_{spec}$, and $\lambda_{corr}$ are non-negative weighting coefficients.

\begin{table*}[htbp]
\centering
\small
\caption{Generative quality (TS-FID $\downarrow$) and downstream utility ($\Delta$Acc $\uparrow$) across three benchmark datasets, reported as mean $\pm$ std over three seeds; best in bold.}
\label{tab:main_results}
\begin{tabular}{lcccccc}
\toprule
\multirow{2}{*}{Method} & \multicolumn{2}{c}{TUEV} & \multicolumn{2}{c}{BCIC-IV-2a} & \multicolumn{2}{c}{SEED-IV} \\
\cmidrule(lr){2-3} \cmidrule(lr){4-5} \cmidrule(lr){6-7}
 & TS-FID $\downarrow$ & $\Delta$Acc $\uparrow$ & TS-FID $\downarrow$ & $\Delta$Acc $\uparrow$ & TS-FID $\downarrow$ & $\Delta$Acc $\uparrow$ \\
\midrule
ESC-GAN          & $612.45\pm18.32$ & $1.24\pm0.42\%$ & $845.20\pm22.71$ & $0.13\pm0.56\%$ & $1102.67\pm28.94$ & $0.85\pm0.38\%$ \\
Diffusion-TS  & $478.36\pm14.07$ & $0.95\pm0.31\%$  & $702.14\pm19.38$ & $0.48\pm0.39\%$ & $893.52\pm23.61$  & $0.42\pm0.27\%$  \\
PaD-TS               & $412.60\pm11.85$ & $1.67\pm0.28\%$  & $635.77\pm16.92$ & $0.15\pm0.33\%$  & $810.29\pm20.15$  & $0.76\pm0.24\%$  \\
JET                   & $360.78\pm9.64$  & $2.81\pm0.25\%$  & $518.89\pm13.47$ & $0.62\pm0.22\%$  & $748.01\pm17.83$  & $1.80\pm0.21\%$  \\
\textbf{Ours}                           & $\mathbf{213.92\pm6.53}$ & $\mathbf{9.58\pm0.47\%}$ & $\mathbf{196.32\pm7.85}$ & $\mathbf{2.58\pm0.29\%}$ & $\mathbf{437.84\pm11.26}$ & $\mathbf{2.61\pm0.33\%}$ \\
\bottomrule
\end{tabular}
\end{table*}

\section{Experiments}

\subsection{Datasets and Preprocessing}

To comprehensively evaluate the proposed framework, we conduct experiments on three publicly available EEG datasets that collectively span clinical event detection, motor imagery decoding, and emotion recognition, namely TUEV \citep{obeid2016tuh,golmohammadi2019automatic}, BCIC-IV-2a \citep{brunner2008bciciv2a,tangermann2012review}, and SEED-IV \citep{zheng2019emotionmeter}.
These datasets differ substantially in channel montage, sampling rate, recording protocol, and label semantics, which allows us to examine whether the position dependent heterogeneity is a general property of EEG signals rather than an artifact specific to a single acquisition setting \citep{yang2023biot,jiang2024labram}.
TUEV comprises 16 channel event-centered windows under a subject-disjoint split; BCIC-IV-2a comprises 22 channel windows aligned to the motor imagery period under a cross-session split; and SEED-IV comprises 62 channel windows under a trial-level split, each chosen to match the dataset's structure and standard evaluation practice and to avoid information leakage between training, validation, and test sets.

Across all three datasets, the patchified representation uses a fixed patch length of 200 time points, so only the channel count $C$ and patch count $N$ vary across benchmarks, while the tokenization scheme remains unchanged \citep{yang2023biot,jiang2024labram,wang2026jet}.
This unified representation lets us apply the proposed framework across datasets with different montages and task semantics, without dataset specific architectural changes.
For the downstream augmentation experiments in Section~\ref{sec:main_results}, synthetic samples for each method are generated to match the per-class sample count of the real training split, so that augmentation changes the quality of samples available to the classifier without altering the underlying class balance; classifier training and generation hyperparameters are detailed in Section~\ref{sec:metrics}.

\subsection{Baselines}

We compare our proposed framework against four representative EEG or general time series generation methods spanning GAN based, diffusion based, and flow matching based paradigms.
\textbf{ESC-GAN} \citep{zhang2025escgan} is a GAN based EEG editing framework that transforms reference signals into under-represented emotional subspaces under diversity aware and boundary aware losses.
\textbf{Diffusion-TS} \citep{yuan2024diffusionts} and \textbf{PaD-TS} \citep{li2024padts} are diffusion based time series generators, the former reconstructing the clean sample at each step under disentangled temporal representations and a Fourier based loss, the latter additionally preserving population level statistical properties such as cross dimensional correlation through a dual channel encoder architecture.
\textbf{JET} \citep{wang2026jet} reformulates EEG generation as conditional flow matching with principled spectral, temporal, and statistical constraints, but still assumes a single global time progression shared across all channels and time segments, making it the strongest and most directly comparable baseline to our work.
All baselines are reimplemented following their original papers and trained on the same dataset splits, with model selection on the validation set and comparisons reported on the held out test set.

\subsection{Metrics}
\label{sec:metrics}

\begin{figure*}[t]
\centering
\includegraphics[width=\textwidth]{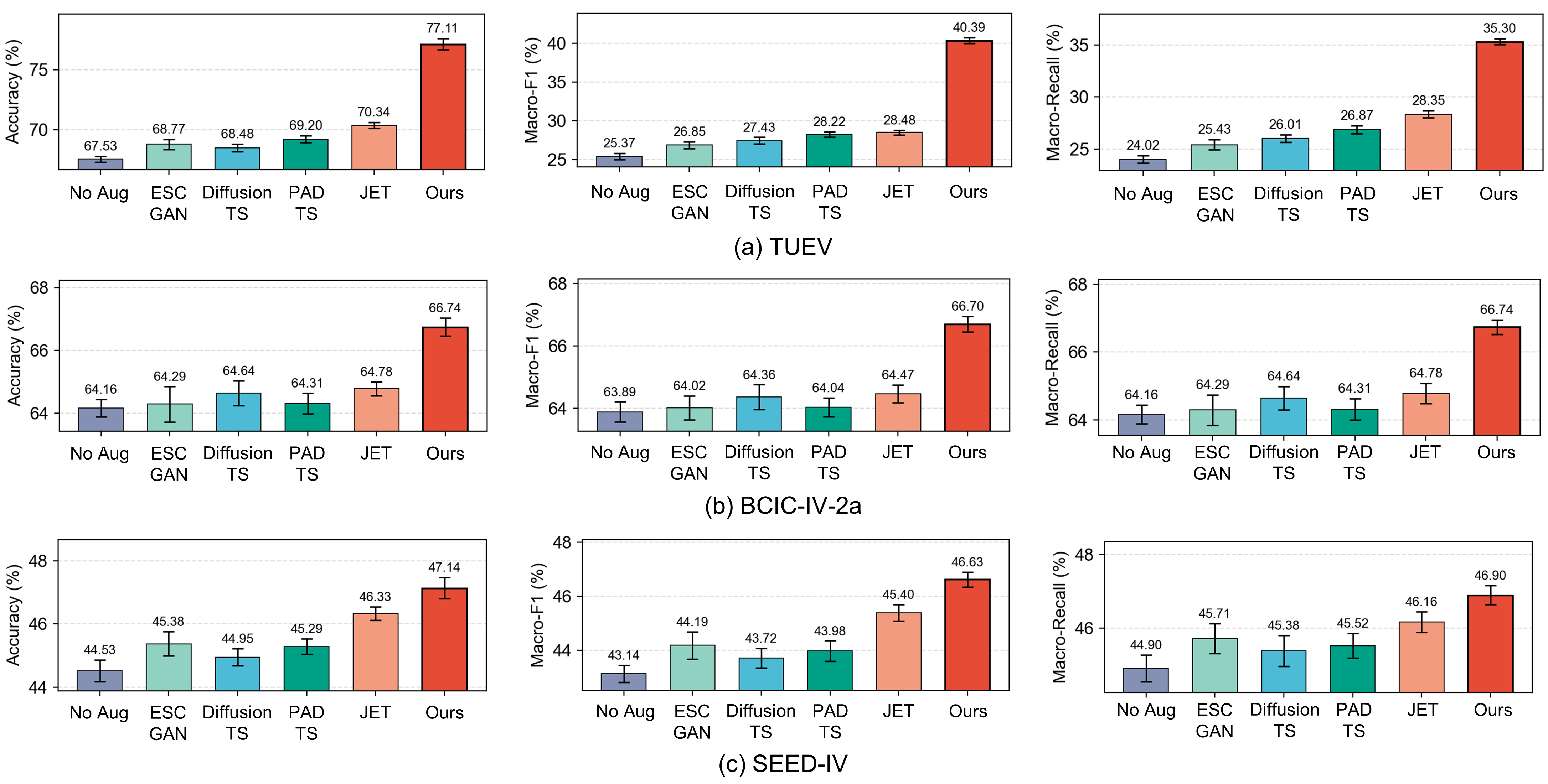}
\caption{Downstream classification performance (Accuracy, Macro-F1, Macro-Recall) on (a) TUEV, (b) BCIC-IV-2a, and (c) SEED-IV: real data only (No Aug) vs. real data augmented by each method. Error bars denote std over three seeds.}
\label{fig:downstream}
\end{figure*}

To provide a holistic assessment of generation quality, we adopt a two dimensional evaluation protocol covering distributional fidelity and downstream clinical utility, following prior practice in EEG and general time series generation \citep{wang2026jet}.
We report a domain specific Time-Series Fréchet Inception Distance (TS-FID), which measures the Fréchet distance between real and generated signals in a compact spectral feature space, since the standard FID relies on a vision pretrained extractor not applicable to EEG \citep{heusel2017gans}.
We further evaluate downstream utility by augmenting the real training set with synthetic samples and measuring the resulting accuracy gain of a downstream EEG classifier, since high spectral fidelity does not necessarily translate into discriminative features useful for classification \citep{fahimi2021ganbci}.
For Table~\ref{tab:main_results} and Figure~\ref{fig:downstream}, the evaluation classifier is EEGNet \citep{lawhern2018eegnet}, a lightweight convolutional architecture comprising a temporal convolution block, a depthwise spatial convolution across channels, and a separable convolution for feature aggregation, followed by a linear classification head, trained from scratch under identical hyperparameters across all augmentation methods.
For Table~\ref{tab:cbramod_delta_acc}, we additionally evaluate with CBraMod \citep{wang2025cbramod} as a large-scale pretrained evaluation backbone, consistent with common practice in prior work.

\begin{table*}[t]
\centering
\small
\caption{Downstream utility ($\Delta$Acc) using CBraMod as the evaluation classifier, with 1$\times$ generated data mixed into the real training set. Mean $\pm$ std over three seeds; best in bold.}
\label{tab:cbramod_delta_acc}
\begin{tabular}{lccc}
\toprule
Method & TUEV & BCIC-IV-2a & SEED-IV \\
\midrule
ESC-GAN \citep{zhang2025escgan}          & $0.25\pm0.31\%$ & $0.12\pm0.38\%$ & $0.26\pm0.29\%$ \\
Diffusion-TS \citep{yuan2024diffusionts} & $0.42\pm0.27\%$  & $0.35\pm0.33\%$ & $0.28\pm0.25\%$  \\
PaD-TS \citep{li2024padts}               & $0.98\pm0.24\%$  & $0.68\pm0.29\%$  & $0.87\pm0.23\%$  \\
JET \citep{wang2026jet}                  & $1.64\pm0.22\%$  & $1.53\pm0.26\%$  & $1.41\pm0.21\%$  \\
\textbf{Ours}                           & $\mathbf{2.39\pm0.19\%}$ & $\mathbf{2.74\pm0.24\%}$ & $\mathbf{2.15\pm0.20\%}$ \\
\bottomrule
\end{tabular}
\end{table*}

\subsection{Main Results}
\label{sec:main_results}

Table~\ref{tab:main_results} reports the comparison between our proposed framework and four representative baselines across TUEV, BCIC-IV-2a, and SEED-IV.
Across all three datasets, which differ substantially in channel count, class number, and task semantics, our framework consistently achieves the lowest TS-FID and the highest $\Delta$Acc, with GAN based and diffusion based methods offering only limited gains and the flow matching based JET remaining the strongest baseline, consistent with the progression from discrete to continuous generative formulations discussed in Related Work.
Relative to JET, our framework reduces TS-FID by 40.7\%, 62.2\%, and 41.5\% on TUEV, BCIC-IV-2a, and SEED-IV respectively, and improves $\Delta$Acc by 6.77, 1.96, and 0.81 percentage points on the same three datasets, with the largest gains of 62.2\% and 6.77 percentage points (a 3.4$\times$ relative gain) achieved on BCIC-IV-2a and TUEV respectively, confirming that explicitly modeling the heterogeneous reconstruction difficulty across channels and time positions yields synthetic samples that translate more consistently into downstream gains.

Table~\ref{tab:cbramod_delta_acc} further evaluates downstream utility using CBraMod \citep{wang2025cbramod}, a large-scale pretrained EEG foundation model, in place of the lightweight evaluation classifier used in Table~\ref{tab:main_results}.
Our method again achieves the largest accuracy gain on all three datasets, exceeding the strongest baseline JET by 0.75, 1.21, and 0.74 percentage points on TUEV, BCIC-IV-2a, and SEED-IV respectively, with the largest margin of 1.21 percentage points achieved on BCIC-IV-2a, reinforcing that not all EEG moments are equal even under a large-scale pretrained backbone.

Figure~\ref{fig:downstream} presents the downstream classification performance under each augmentation method, using the lightweight decoding model described in Section~\ref{sec:metrics}.
Our method consistently achieves the best classification performance on all three datasets, with the largest margin on TUEV, where it raises accuracy from 67.53\% to 77.11\% while improving Macro-F1 by 1.59$\times$ and Macro-Recall by 1.47$\times$ over the No Aug baseline, and a smaller but consistent margin on the more class balanced BCIC-IV-2a and SEED-IV.

\begin{table}[t]
\centering
\small
\caption{Leave-one-out ablation on TUEV. Mean $\pm$ standard deviation over three random seeds.}
\label{tab:ablation}
\begin{tabular}{lcc}
\toprule
Configuration & TS-FID $\downarrow$ & $\Delta$Acc $\uparrow$ \\
\midrule
Baseline        & $360.78\pm9.64$ & $2.81\pm0.25\%$ \\
w/o Factorized Attn            & $241.87\pm7.15$ & $6.47\pm0.31\%$ \\
w/o PATS                       & $268.53\pm7.92$ & $5.12\pm0.34\%$ \\
w/o Spectral Loss              & $226.35\pm6.88$ & $8.03\pm0.28\%$ \\
Full Model              & $\mathbf{213.92\pm6.53}$ & $\mathbf{9.58\pm0.47\%}$ \\
\bottomrule
\end{tabular}
\end{table}

\begin{table}[t]
\centering
\small
\caption{Loss design ablation on TUEV. Mean $\pm$ standard deviation over three random seeds.}
\label{tab:loss_ablation}
\begin{tabular}{lcc}
\toprule
Configuration & TS-FID $\downarrow$ & $\Delta$Acc $\uparrow$ \\
\midrule
w/o Reweight ($\alpha{=}0$) & $219.47\pm7.02$ & $8.41\pm0.36\%$ \\
Single-Res. ($R{=}1$)       & $221.83\pm7.28$ & $8.07\pm0.33\%$ \\
w/o Weighted Recon.         & $217.35\pm6.94$ & $8.72\pm0.31\%$ \\
Full Model (Ours)           & $\mathbf{213.92\pm6.53}$ & $\mathbf{9.58\pm0.47\%}$ \\
\bottomrule
\end{tabular}
\end{table}

\subsection{Ablation Study}

To isolate the contribution of each component, we conduct a leave-one-out ablation study on TUEV, removing Factorized Spatio-Temporal Attention, Position-Adaptive Time Scheduling, and the frequency aligned spectral consistency loss independently from the full model.
Table~\ref{tab:ablation} shows that removing any single component degrades both TS-FID and $\Delta$Acc while every ablated configuration still clearly outperforms JET, with PATS contributing the largest share of the improvement, confirming that the three components are individually necessary and jointly complementary.

To verify that the frequency aligned spectral consistency loss benefits from each of its specific design choices rather than acting as a single monolithic term, we further disable the frequency reweighting ($\alpha=0$), replace the multi-resolution STFT with a single resolution, and disable the difficulty weighted reconstruction term ($\gamma=0$) independently on top of the full model.
Table~\ref{tab:loss_ablation} shows that all three choices contribute positively, though with a considerably smaller degradation than removing the entire spectral loss in Table~\ref{tab:ablation}, with frequency reweighting contributing the most, consistent with its role in compensating for the power law spectral bias of EEG, followed by the multi-resolution design, which captures both transient and quasi-stationary structure, and the difficulty weighted reconstruction term, which acts as a complementary refinement on top of PATS.

\subsection{Visualization}

\begin{figure}[t]
\centering
\includegraphics[width=0.5\linewidth]{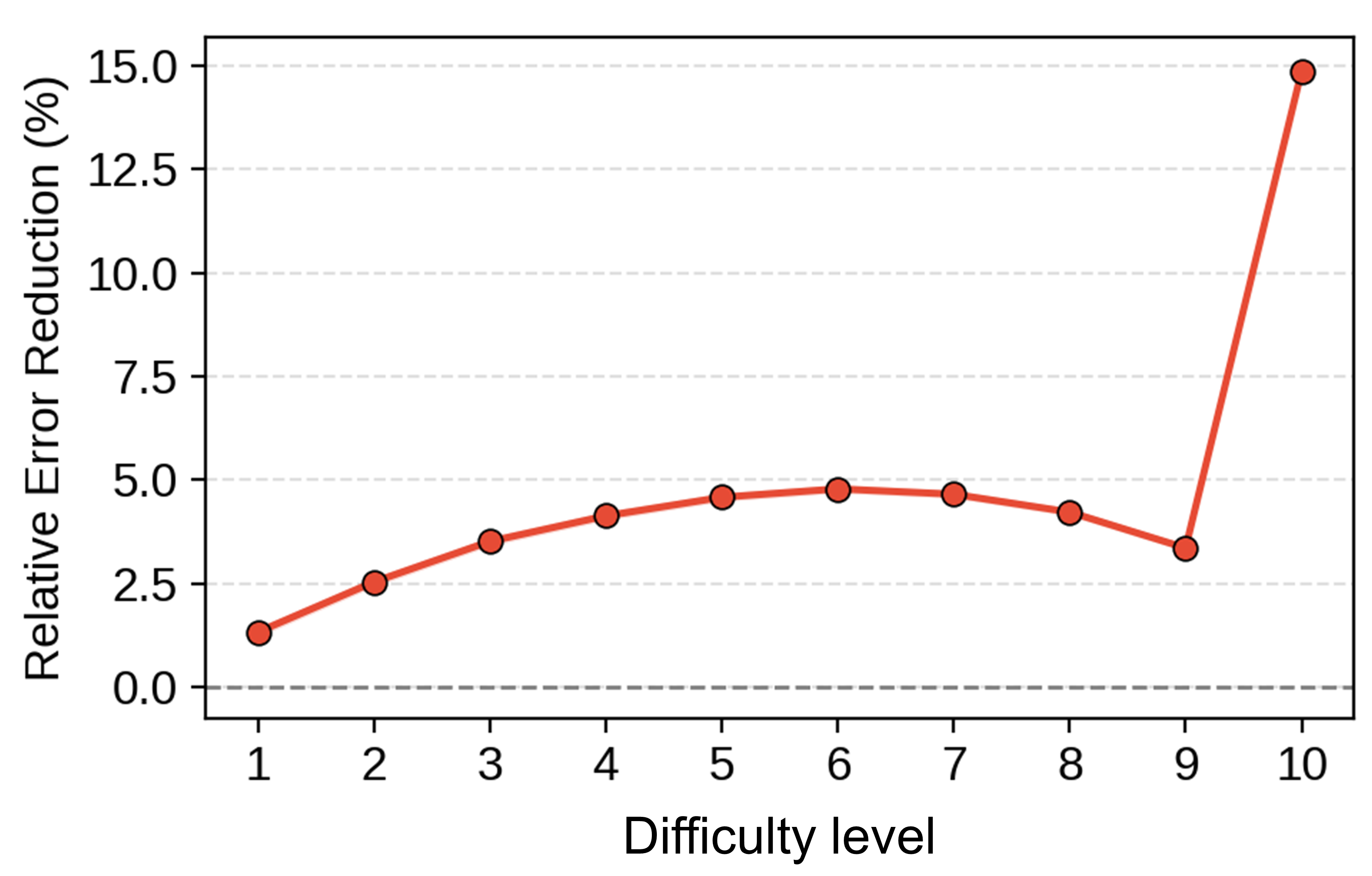}
\caption{Difficulty-stratified relative error reduction on TUEV. Instances are ranked by baseline (JET) error and grouped into ten equal-count deciles from easiest (1) to hardest (10); each point shows the mean relative error reduction of our framework over the baseline.}
\label{fig:difficulty_curve}
\end{figure}

Figure~\ref{fig:difficulty_curve} shows that the relative error reduction achieved by our framework is not uniform across difficulty levels, but concentrates disproportionately on the hardest positions.
The improvement rises steadily from decile 1 to decile 6, from 1.3\% to 4.8\%, then plateaus and slightly declines through deciles 7 to 9, from 4.7\% to 3.3\%.
At decile 10, relative error reduction jumps sharply to 14.9\%, nearly three times the level of any other decile.
Since the saliency map $\mathbf{s}$ is a position-wise statistic shared across the training set rather than recomputed per instance, this analysis examines whether a schedule derived from dataset-level difficulty yields disproportionate benefits on the positions a fixed baseline finds hardest, rather than establishing instance-level content sensitivity.
This sharp, isolated jump at the extreme tail supports our central hypothesis that a small subset of channel-time positions, corresponding to recurring transient events and pathological discharges at particular positions, are disproportionately difficult under a uniform time schedule, and that position adaptive scheduling yields its largest benefit precisely by identifying and prioritizing these positions rather than improving uniformly across difficulty levels.

\section{Conclusion}

In this work, we identify that existing flow matching based EEG generation methods assume a single global time progression, overlooking the heterogeneous reconstruction difficulty across channels and time inherent to EEG signals.
To address this, we propose a framework built on Position-Adaptive Time Scheduling, which modulates a position specific time progress based on per position reconstruction error while preserving the endpoints of the flow matching trajectory, a Factorized Spatio-Temporal Attention module that resolves this condition at the token level, and a frequency aligned spectral consistency loss that compensates for the power law spectral bias of EEG.
Extensive experiments on three EEG datasets show that our framework consistently outperforms the strongest prior baseline, reducing TS-FID by up to 62.2\% and improving downstream classification accuracy gain by up to 6.77 percentage points.
Future work may explore instance-level difficulty estimation that adapts to the content of each sample rather than relying solely on dataset-level statistics, extend position adaptive scheduling to multimodal and cross-subject settings, and investigate its applicability to other physiological time series.

\bibliographystyle{plainnat}
\bibliography{references}

\end{document}